\documentclass[%
 aip,
 amsmath,amssymb,
 reprint,%
]{revtex4-2}

\usepackage{graphicx}% Include figure files
\usepackage{dcolumn}% Align table columns on decimal point
\usepackage{bm}% bold math
\usepackage{physics}
\usepackage{caption}
\usepackage{subcaption}
\usepackage{tabularx}
\usepackage[utf8]{inputenc}
\usepackage[T1]{fontenc}
\usepackage{mathptmx}
\usepackage{multirow}
\usepackage{dsfont} % For identity operator symbol
\newcommand\Tstrut{\rule{0pt}{2.6ex}}         % = `top' strut
\usepackage{xcolor}

\begin{document}

\preprint{AIP/123-QED}
\title[
Ethylene photoisomerization and dissociation from LSC dynamics]{Nonadiabatic simulations of photoisomerization and dissociation in ethylene using \textit{ab initio} classical trajectories}
%\title[On-the-fly ab initio nonadiabatic dynamics with variants of LSC-IVR]{On-the-fly ab initio nonadiabatic dynamics with variants of linearized semiclassical initial value representation: Application to ethylene photoisomerization and dissociation}
% Force line breaks with \\

\author{K. Miyazaki}
% \altaffiliation[Also at ]{Physics Department, XYZ University.}%Lines break automatically or can be forced with \\
\author{N. Ananth}%
 \email{ananth@cornell.edu.}
\affiliation{ 
Department of Chemistry and Chemical Biology, Cornell University, Ithaca, New York, 14853, United States%\\This line break forced with \textbackslash\textbackslash
}%

\date{\today}% It is always \today, today,
             %  but any date may be explicitly specified

\begin{abstract}
We simulate the nonadiabatic dynamics of 
photo-induced isomerization and dissociation
in ethylene using \textit{ab initio} classical trajectories in an extended phase space of 
nuclear and electronic variables. This is 
achieved by employing the Linearized Semiclassical Initial 
Value Representation (LSC-IVR) method for 
nonadiabatic dynamics where discrete electronic states are 
mapped to continuous classical variables using either the Meyer-Miller
Stock-Thoss representation or a more recently introduced spin 
mapping approach. 
Trajectory initial conditions are sampled by constraining
electronic state variables to a single initial
excited state, and by drawing nuclear phase space
configurations from a Wigner distribution at 
finite temperature. An ensemble of 
classical \textit{ab initio} trajectories are then generated 
to compute thermal population correlation functions and to 
analyze the mechanisms of isomerization and dissociation. 
Our results serve as a demonstration that 
this parameter-free semiclassical approach 
is computationally efficient and accurate, identifying
mechanistic pathways in agreement with previous theoretical
studies, and also uncovering dissociation pathways observed
experimentally.
\end{abstract}

\maketitle

\section{\label{sec:level1}Introduction}
Nonadiabatic effects that result from the 
coupling of nuclear motion to electronic transitions are 
crucial to understanding the mechanism of various 
interesting photophysical and photochemical phenomena 
including internal conversion, intersystem crossing,  and photodissociation.~\cite{Bircher2017}
Unfortunately, the high dimensionality of complex molecular systems makes 
real-time dynamic simulations on a pre-computed potential energy surface 
challenging and motivates the development of on-the-fly dynamic methods.

The search for computationally efficient and accurate on-the-fly quantum 
dynamic simulations remain an outstanding challenge, particularly in systems 
where nonadiabatic effects play a central role. Although several methods 
have been developed for the simulation of nonadiabatic processes, 
only a handful lend themselves to on-the-fly implementations that 
require not just routine force calculations but also evaluation of 
the nonadiabatic coupling vector at each time step. 
Multiconfigurational Time-dependent Hartree (MCTDH) is, arguably, 
the most exact of these techniques relying on a weighted sum of 
Gaussian nuclear functions centered on a grid while solving the 
Schr\"odinger equation exactly for the electronic
degrees of freedom.~\cite{Richings2018} 
%Existing on-the-fly nonadiabatic methods can be classified 
%by their treatment of nuclear dynamics. Methods 
%like multiconfigurational time-dependent Hartree (MCTDH),~\cite{Meyer1990,Beck2000} variational Gaussian %dynamics,~\cite{Worth2008} and multiple spawning~\cite{Martinez1996} 
%assign a wavepacket to each nuclear degree of freedom (dof).
%On-the-fly MCTDH dynamics was implemented only recently using 
%an approach that employs a weighted sum of Gaussian nuclear 
%functions centered on a grid.~\cite{Richings2018}
Variational Gaussian methods are extensions of MCTDH 
where the nuclear wavefunction is replaced by a Gaussian wavepacket.~\cite{Worth2008}
In multiple spawning and related methods,~\cite{Martinez1996,Shalashilin2014} 
the number of nuclear basis functions increase through `spawning' 
events that occur when electronic states become near-degenerate, 
mimicking the quantum dynamical bifurcation of wavepackets.
Formally, multiple spawning can converge to exact quantum dynamics 
if a sufficiently large number of wavepackets are included, 
and the transition matrix elements are evaluated exactly at each step.~\cite{BenNun2002} 
However, in practice, it is necessary to limit the number of wavepackets spawned 
and approximate matrix elements as is done in approaches like \textit{ab initio} 
multiple spawning (AIMS).~\cite{Worth2008,fedorov2017,Martinez2018}

Mixed quantum-classical methods like surface 
hopping and Ehrenfest dynamics further approximate nuclear motion 
by treating it classically.
In surface hopping~\cite{Tully1990,Tully1998} nuclear wavepackets with finite widths 
in the nuclear coordinates are replaced by independent classical trajectories. 
Each trajectory is allowed to `hop' between surfaces according to a 
stochastic algorithm based on the amplitudes of each electronic state.
Ehrenfest is a mean-field dynamics, where the potential energy for classical 
nuclear propagation is obtained by the weighted average of electronic state energies 
obtained at an instantaneous nuclear geometry.~\cite{Dolt2002} 
While the non-interacting classical trajectories simplify the dynamics,
both surface hopping and Ehrenfest are known to suffer from 
strong coherence and slow decoherence, necessitating additional 
\textit{ad hoc} modifications.~\cite{Bitt1995,Schwartz1996,Prez1997,Volo2000,Hack2001,Bedard2005,Jain2016}

The methods discussed thus far focus on approximately time evolving wavepackets
that can then be used to compute real-time correlation functions necessary to compare
with experimental observables. Semiclassical (SC) and path-integral based methods~\cite{Habershon2013,Makri2022,Ananth2022} 
take an alternate approach, directly approximating the real-time quantum correlation
function. As reviewed recently, there is a hierarchy of SC methods that include 
quantum effects and more specifically nonadiabatic effects to differing extents.~\cite{Shreyas2022} However, only the more classical-limit SC 
methods are sufficiently efficient to allow for on-the-fly dynamic simulations. 
In particular, the Linearized Semiclassical Initial Value Representation (LSC-IVR) 
approach~\cite{Shi2003,Miller2001} is a classical-limit method that 
approximates the quantum correlation functions between two operators, $\hat A$ and 
$\hat B$, as a single phase space integral over a product of Wigner functions evaluated 
at time zero and classically time-evolved phase space configurations at later time $t$
respectively. LSC-IVR is exact at time zero and for harmonic potentials, and offers 
a simple implementation. Recently, the symmetrical quasiclassical (SQC) method 
has been introduced~\cite{Cotton2013_1,Cotton2013_2,Cotton2019}
where the Wigner functions in LSC-IVR are replaced by 
`window' functions that have been successfully 
used to study a range of model chemical systems,~\cite{Cotton2013_2,Cotton2014,Cotton2019, Talbot2022}
\textit{ab initio} dynamics using the so-called quasi-diabatic electronic states~\cite{Frank2019,Frank2021} as well 
as using adiabatic electronic states with an approximate 
integration scheme.~\cite{Talbot2022}
Unfortunately, unlike LSC-IVR, 
SQC requires the choice of an \textit{ad hoc} window function that can significantly
affect simulation accuracy.~\cite{Subot16, Frank2019,Frank2021} 

In this paper, we implement a first on-the-fly nonadiabatic dynamic simulation 
with LSC-IVR using \textit{ab initio} classical trajectories in an extended phase space 
of nuclear and adiabatic electronic state variables.
The classical Hamiltonian employed by these semiclassical methods is obtained 
by mapping discrete electronic state variables to continuous Cartesian variables.
We explore two mapping protocols: the well-established Meyer-Miller-Stock-Thoss (MMST) approach~\cite{stock97,Thoss99,Meyer1990,Miller2016} and the more recently introduced 
spin mapping that shows great promise in model system studies.~\cite{Runeson2019,Runeson2020} 
To enable on-the-fly simulations, we use an adiabatic electronic state representation~\cite{Meyer1990,Ananth2007,Cotton17}
in conjunction with the `kinematic' momentum integration scheme.~\cite{Cotton17} 
Trajectory initial conditions for the nuclei are sampled from an initial Wigner
distribution, while electronic state variables are sampled to ensure that only
a single excited electronic state is populated at time zero. We calculate the 
real-time population correlation function by generating classical trajectories
from the mapping Hamiltonian, with the potential gradients, energies,
and the nonadiabatic coupling vector calculated at each time step from a CASSCF(2o,2e) 
electronic structure calculation. We then analyze the resulting 
ensemble of trajectories to identify the mechanisms of photoisomerization and 
dissociation pathways in ethylene, and compare our results against previous 
theoretical and experimental studies. We conclude with a detailed discussion
highlighting the significant advantages of the LSC-IVR approach for on-the-fly nonadiabatic 
simulations while also outlining outstanding challenges.

\section{\label{sec:theory}Theory}
\subsection{\label{sec:LSC-IVR}LSC-IVR approximation to quantum correlation functions}
In the path integral representation of quantum mechanics, the real-time quantum 
correlation function, 
\begin{align}
\label{eq:QCF}
C_{AB}(t) = \Tr\left[ \hat{A}e^{i\hat{H}t/\hbar}\hat{B}e^{-i\hat{H}t/\hbar}\right],
\end{align}
is expressed as a double sum over all possible forward and backward paths in coordinate
space with an overall phase corresponding to the difference in action between the 
forward and backward paths. Truncating the difference in action to first order yields
the LSC-IVR approximation~\cite{Shi2003,Sun1998} 
to the correlation function in Eq.~\eqref{eq:QCF},
\begin{align}
\label{eq:LSC_CF}
C_{AB}^{LSC}(t) = \cfrac{1}{(2\pi\hbar)^{N}} \int d\mathbf{X}_{0}\int d\mathbf{P}_{0} \, A_{\mathcal{W}}(\mathbf{X}_{0},\mathbf{P}_{0})B_{\mathcal{W}}(\mathbf{X}_{t},\mathbf{P}_{t}).
\end{align}
In Eq.~\eqref{eq:LSC_CF}, $O_{\mathcal{W}}(\mathbf{X},\mathbf{P})$ represents 
the Wigner transform of $\hat{O}$ defined as
\begin{align}
\label{eq:wigner_transform}
O_{\mathcal{W}}(\mathbf{X},\mathbf{P}) = %\cfrac{1}{(2\pi\hbar)^{N}} 
\int d\bm{\Delta} \mel{\mathbf{X}+\frac{\bm{\Delta}}{2}}{\hat{O}}{\mathbf{X}-\frac{\bm{\Delta}}{2}} e^{-i\mathbf{P}\bm{\Delta}/\hbar},
\end{align}
where $N$ is the number of system degrees of freedom, 
and $(\mathbf{X},\mathbf{P})$ are phase space vectors.
An ensemble of trajectories are generated by sampling initial conditions from $A_{\mathcal{W}}(\mathbf{X}_{0},\mathbf{P}_{0})$ and propagated for time $t$ according to classical equations of motion generated by the Hamiltonian, $H(\mathbf{X},\mathbf{P})$. The function $B_\mathcal{W}$ is then evaluated at the time-evolved phase space variables $(\mathbf{X}_{t},\mathbf{P}_{t})$.
Calculating the LSC-IVR correlation function in Eq.~\ref{eq:LSC_CF} is generally efficient, incorporating quantum effects like nuclear tunneling and zero-point energy at a computational cost similar to a classical simulation. 

\subsection{\label{sec:MM-ST}{Thermal Correlation Functions and Nonadiabatic Dynamics}}
The LSC-IVR expression for a real-time thermal correlation function requires 
the Wigner transform of the density operator,
$\hat{A}\equiv \hat{\rho}=e^{-\beta \hat H}$ (Eq.~\ref{eq:wigner_transform}), where $\beta=1/k_B T$.
For a multi-state system, we assume the initial density operator is separable, 
$\hat{\rho}=\hat{\rho}_{e} \hat{\rho}_{n}$, where $\hat{\rho}_{e}$ and $\hat{\rho}_{n}$ are 
the electronic and nuclear density operators, respectively, 
and $\Tr_{e}\left[\hat{\rho}_{e}\right] = \Tr_{n}\left[\hat{\rho}_{n}\right] = 1$. 
Under this assumption, the Wigner transform of $\hat{A}$ can be written as the 
product of separate transforms: $A_{\mathcal{W}}(\mathbf{R},\mathbf{P},\mathbf{x},\mathbf{p}) = 
\left[\hat{\rho}_{e}\right]_{\mathcal{W}} (\mathbf{x},\mathbf{p}) 
\left[\hat{\rho}_{n}\right]_{\mathcal{W}} (\mathbf{R},\mathbf{P})$ where $\left(\mathbf{R},\,\mathbf{P}\right)$ are nuclear phase space vectors and the $\left(\mathbf{x},\,\mathbf{p}\right)$ are the electronic state phase space variables. 
Assuming a thermal distribution of harmonic normal modes, it is possible to express
the Wigner transform of nuclear density as 
\begin{align}
\label{eq:nuc_density}
\left[\hat{\rho}_{n}\right]_{\mathcal{W}} 
    &= \left[ e^{-\beta \hat{H}_{n}}\right]_{\mathcal{W}} \notag \\
    &= \prod_{i=1}^{N-F}\cfrac{1}{2\pi} \, 
    \exp\left[ -\tanh\left(\cfrac{\beta \omega_{i}}{2}\right) \left( \cfrac{1}{\mu_{i}\omega_{i}}P_{i}^{2} + \mu_{i}\omega_{i}R_{i}^{2}\right)\right],
\end{align}
where $\mu_{i}$ and $\omega_{i}$ are the reduced mass and frequency of the $i$-th vibrational mode, $N$ is 
the total number of degrees of freedom in the system (electronic and nuclear) and $F$ 
is the number of electronic states.

For photo-initiated processes, the electronic density is defined as 
the projection operator onto a single initially occupied $i^\text{th}$ 
electronic state, 
\begin{align}
\left[\hat{\rho}_{e}\right]_{W} = \left[ \ket{i}\bra{i} \right]_{W}.
\label{eq:el_den_proj}
\end{align}
In the SC framework, the discrete state-space electronic density matrix 
and the corresponding multi-state system Hamiltonian are treated by mapping 
them to a continuous Cartesian variable representation, and we describe 
two such schemes below.

\subsubsection{MMST mapping:} 
The MMST approach effectively maps the occupation of an electronic state 
to a single excitation quantum of a corresponding harmonic oscillator,~\cite{stock97}
%such that $\ket{n} \rightarrow \ket{0_{1}...1_{n}...0_{F}}$.~\cite{Meyer79,stock97}
%Accordingly, the electronic density matrix elements are mapped onto singly excited harmonic oscillator basis such that\cite{stock97}
\begin{equation}
\begin{aligned}
\label{eq:mmst_map}
\ket{n}\bra{m} \rightarrow \hat{a}^{\dagger}_{n} \hat{a}_{m}
\end{aligned}
\end{equation}
where $F$ is the total number of electronic states and the creation and 
annihilation operators of $k$th harmonic oscillator 
are $\hat{a}^{\dagger}_{k}=(\hat{x}_{k}-i\hat{p}_{k})/\sqrt{2}$ 
and $\hat{a}_{k}=(\hat{x}_{k}+i\hat{p}_{k})/\sqrt{2}$ in terms of electronic phase space variables.  
Using this mapping, the multi-state Hamiltonian expressed in the adiabatic electronic
state representation can be written as,~\cite{Meyer79,Meyer80,Cotton17}
\begin{align}
\label{eq:mmst_adiabH}
    H 
    &= \frac{(\mathbf{P}+\Delta\mathbf{P})^{2}}{2\mu} + \sum_{n}^{F} \frac{1}{2} \left( p_{n}^{2} + x_{n}^{2} - \gamma \right) E_{n}(\mathbf{R}) \notag \\
    &= \frac{\mathbf{P}^{2}_{\text{kin}}}{2\mu} + \sum_{n}^{F} \frac{1}{2} \left( p_{n}^{2} + x_{n}^{2} - \gamma \right) E_{n}(\mathbf{R}).
\end{align}
We note that the MMST  mapping is exact when $\gamma=1$, and this is the value we use in this manuscript although some semiclassical simulations treat $\gamma$ as a zero-point energy (ZPE) parameter that can be modified to increase numerical stability.
In Eq.~\ref{eq:mmst_adiabH}, $E_{n}(\mathbf{R})$ is the adiabatic 
energy of the $n$-th electronic state, 
the kinematic momentum $\mathbf{P}_{\text{kin}} = \mathbf{P}+\Delta\mathbf{P}$, with 
\begin{align}
\label{eq:DeltaP}
    \Delta\mathbf{P} = \sum_{n \neq m}^{F} x_{n}p_{m}\mathbf{d}_{nm}(\mathbf{R}),
\end{align}
where $\mathbf{d}_{nm}(\mathbf{R}) = \mel{\phi_{n}}{\frac{\partial}{\partial\mathbf{R}}}{\phi_{m}}$ is the nonadiabatic coupling vector between the electronic states $\ket{\phi_{n}}$ and $\ket{\phi_{m}}$.~\cite{Cotton17}
%The transition from the position and momentum operators to the classical phase space variables in Eq.~\ref{eq:mmst_adiabH} is made through the Wigner transform.
It is further possible to construct an equivalent, exact, symmetrized form of 
the MMST Hamiltonian,
\begin{align}
\label{eq:mmst_adiabH_sym}
H &= \cfrac{\mathbf{P}_{\text{kin}}^{2}}{2\mu} + \cfrac{1}{F}\sum_{n}^{F} E_{n}(\mathbf{R}) \notag \\ 
    & \,\,\,\,\,\,\,\, + \cfrac{1}{F}\sum_{n,m}^{F} \cfrac{1}{4} \, \left(p_{n}^{2}+x_{n}^{2}-p_{m}^{2}-x_{m}^{2}\right)\left(E_{n}(\mathbf{R})-E_{m}(\mathbf{R})\right) \notag \\
    &= \cfrac{\mathbf{P}_{\text{kin}}^{2}}{2\mu} + V_{\text{eff}},
\end{align}
such that the equations of motion are independent of the ZPE parameter.~\cite{Cotton2013_2,Cotton17} 

In order to evaluate the electronic density matrix as defined 
in Eq.~\ref{eq:el_den_proj}, we consider the Wigner transform of the projection operator.
In the MMST framework, the phase space expression obtained through 
Wigner transformation differs depending on the quantum mechanical 
definition,~\cite{Sun1998,Pierre2017,Saller2019} 
and we consider both forms in this work.
Defining the projection operator in the singly excited oscillator (SEO) basis yields the so-called Wigner 
population estimator,~\cite{Duke2015}
\begin{align}
\label{eq:el_seo_density}
P^i_{\mathcal{W}}
&=\left[ \ket{i}\bra{i} \right]_{\mathcal{W}}^{\text{SEO}} \notag \\
&= 2^{F+1} \left( x_{i}^{2} + p_{i}^{2} - \cfrac{1}{2} \right) \exp \left[ -\sum_{j}^{F}\left( x_{j}^{2} + p_{j}^{2} \right) \right].
\end{align}
Expressing the projection operators using the mapping variable operators yields
the SC population estimator,~\cite{Pierre2017}
\begin{align}
\label{eq:el_density}
P^i_\text{SC}=
\left[\cfrac{1}{2} \left( \hat x_{i}^{2} + \hat p_{i}^{2} - 1 \right)\right]_{\mathcal{W}}
=\cfrac{1}{2} \left( x_{i}^{2} + p_{i}^{2} - 1 \right).
\end{align}

%A key feature of the adiabatic MMST Hamiltonian in Eq.~\ref{eq:mmst_adiabH_sym} is that 
%the classical equations of motion do not involve the second derivative of the nonadiabatic
%coupling vector,
%\begin{subequations}
%\label{eq:EOM}
%\begin{alignat}{2}
%&\Dot{\mathbf{P}}_{\text{kin}}
%    = -\sum_{n,m}^{F} \cfrac{1}{2}\left( p_{n}p_{m}+x_{n}x_{m}\right) \left( E_{n}(\mathbf{R})-E_{m}(\mathbf{R})\right) \mathbf{d}_{nm}(\mathbf{R}) \notag \\
%    &\qquad\quad -\pdv{V_{\text{eff}}}{\mathbf{R}} \\
%&\Dot{\mathbf{R}}
%    = \cfrac{\mathbf{P}_{\text{kin}}}{\mu}\\
%&\Dot{p}_{n}
%    = -x_{n}\cfrac{1}{F} \, \sum_{m}^{F} \left(E_{n}(\mathbf{R})-E_{m}(\mathbf{R})\right) + \sum_{m}^{F} p_{m} \, \mathbf{d}_{mn}(\mathbf{R}) \cdot \cfrac{\mathbf{P}_{\text{kin}}}{\mu} \\
%&\Dot{x}_{n}
%    = p_{n}\cfrac{1}{F} \, \sum_{m}^{F} \left(E_{n}(\mathbf{R})-E_{m}(\mathbf{R})\right) + \sum_{m}^{F} x_{m} \, \mathbf{d}_{mn}(\mathbf{R}) \cdot \cfrac{\mathbf{P}_{\text{kin}}}{\mu} 
%\end{alignat}
%\end{subequations}

\subsubsection{\label{sec:SM}Spin mapping}
%The spin mapping approach maps $F$ electronic states to a $(F-1)/2$ spin coherent state basis.
%The spin-mapping (SM) scheme maps $F$ electronic states to spin-$(F-1)/2$ system. 
As an alternative to MMST mapping, various schemes that map electronic states to spin variables 
have been previously proposed;~\cite{Meyer79spin, Thoss99, Cotton15spin}
here we employ a recently introduced scheme that appears to out-perform MMST mapping 
in model system studies.~\cite{Runeson2019,Runeson2020,Bossion2022}

The spin mapping (SM) approach maps an $F$-level electronic system to an $(F-1)/2$ spin system,~\cite{Runeson2019} 
and the corresponding Hamiltonian is obtained by recognizing that the $F$-level system Hamiltonian
can be exactly expressed as a linear combination of the $F^2-1$ spin angular momentum matrices
that define a spin $(F-1)/2$ system and the identity operator,
\begin{align}
\label{eq:spinH1}
\hat{H} = H_{0}\hat{\mathds{1}} + \sum_{i=1}^{F^{2}-1}H_{i}\hat{S}_{i}
\end{align}
where $H_0$ and the $\{H_{i}\}$ are the expansion coefficients,
and the spin angular momentum matrices, 
$\lbrace \hat{S}_i\rbrace_{i=1}^{F^{2}-1}$, are traceless and orthogonal.
The Hamiltonian in Eq.~\eqref{eq:spinH1} can be written in terms of 
continuous phase-space Cartesian operators using the Stratonovich-Weyl
transforms.~\cite{strato1957,Brif1998,Brif1999}
Interestingly, after symmetrization, in the adiabatic electronic 
state representation the resulting SM Hamiltonian is identical to 
the symmetrized MMST Hamiltonian in Eq.~\eqref{eq:mmst_adiabH_sym}. 
In this manuscript, we focus on the 
$W$-representation of the Stratonovich-Weyl transform, 
which results in a correlation function analogous to that of LSC-IVR (Eq.~\ref{eq:LSC_CF}),
%most closely resembles the Wignertransformations in LSC-IVR 
and that recent work suggests is the most successful choice 
of representation.~\cite{Runeson2019, Runeson2020}

As before, we define the initial electronic density matrix as a projection
onto a single discrete electronic states. Since the specific form of the projection
operator depends on the number of electronic states, we consider the case
where $F=3$ in line with our study of the photoisomerization of ethylene. 
Expressing the projection operator in terms of spin matrices,
\begin{align}
\label{eq:elec_proj_expansion}
\ket{j}\bra{j} = C_{0}\hat{\mathds{1}} + \sum_{i}^{F^{2}-1}C_{i}\hat{S}_{i},
\end{align}
and using the $W$-representation leads to expressions
for the population estimators,~\cite{Runeson2020}
\begin{subequations}
\label{eq:3_elec_proj_cart}
\begin{align}
P_{\text{SM}}^1=
\left[ \ket{1}\bra{1} \right]_{\text{SM}} 
&= \cfrac{1}{3} + \cfrac{1}{6} \left( 2 r_{1}^{2} - r_{2}^{2} - r_{3}^{2} \right) \\
P_{\text{SM}}^2=
\left[ \ket{2}\bra{2} \right]_{\text{SM}} 
&= \cfrac{1}{3} - \cfrac{1}{6} \left( r_{1}^{2} - 2 r_{2}^{2} + r_{3}^{2} \right) \\
P_{\text{SM}}^3=
\left[ \ket{3}\bra{3} \right]_{\text{SM}} 
&= \cfrac{1}{3} - \cfrac{1}{6} \left( r_{1}^{2} + r_{2}^{2} - 2 r_{3}^{2} \right)  
\end{align}
\end{subequations}
where $r_{i}^{2} = x_{i}^{2} + p_{i}^{2}$ and $\sum_{j}^{F} \left[ \ket{j}\bra{j} \right]_{\text{SM}} = 1$.

\begin{table}
\setlength{\tabcolsep}{18pt}
\caption{\label{tab:samp_func_radii} The radii of the sampling functions for the 
three electronic population estimators ($\mathcal{W}$, SC, and $\text{SM}$) for 
a system with three coupled electronic states.}
\begin{ruledtabular}
\begin{tabular}{ccc}
    \multirow{2}{*}{$M$ (Eq.~\ref{eq:wig_sampling})} & \multicolumn{2}{c}{Sampling radius, $r_{k}$}\\
\cline{2-3}\Tstrut
 & \text{Occupied} & \text{Unoccupied} \\
\hline\Tstrut
$\mathcal{W}$ (Eq.~\ref{eq:el_seo_density}) & 1.55892\footnote{The larger of the two roots is chosen following Ref.~\onlinecite{Duke2015}.} & 1/2\\
\Tstrut
$\text{SC}$ (Eq.~\ref{eq:el_density}) & 3 & 1 \\
\Tstrut
$\text{SM}$ (Eq.~\ref{eq:3_elec_proj_cart}) & 8/3 & 2/3\\
\end{tabular}
\end{ruledtabular}
\end{table}

\subsection{\label{sec:simulation_detail}Simulation details}
Excited state dynamics in the LSC-IVR framework are obtained from the 
population correlation function,
\begin{align}
    \nonumber
    C_{P_j}(t)&=\frac{1}{(2\pi\hbar)^N} 
    \int d{\bf R}_{0} \int d{\bf P}_{0} \int d{\bf x}_{0} \int d{\bf p}_{0} \\
    & \times 
    \left[ \hat \rho_n \right]_\mathcal{W}({\bf R}_0, {\bf P}_0)
    \left[ \hat \rho_e \right]_M({\bf x}_0, {\bf p}_0) P^j_M\left( {\bf x}_t, {\bf p}_t \right),
    \label{eq:lsc_pop_cf}
\end{align}
where the Wigner transform of the nuclear density matrix, 
$\left[ \hat \rho_n \right]_\mathcal{W}$, is defined in Eq.~\eqref{eq:nuc_density}
and used to sample the initial nuclear phase space variables 
at temperature $T = 300$K, with the frequencies and the reduced masses 
calculated at the equilibrium geometry of the $S_0$ state. 
For a system initially in a single excited state, the 
electronic density matrix in Eq.~\eqref{eq:lsc_pop_cf} can be 
expressed in terms of electronic population estimators,
\begin{align}
\label{eq:wig_sampling}
\left[\hat{\rho}_{e}\right]_{M} 
= \delta \left(P_{M}^1-1\right)\prod_{j\neq 1}^{2} \delta \left(P_{M}^j \right),
\end{align}
where the subscript $M=\mathcal{W},\,\text{SC},\,\text{SM}$ indicates
the specific choice of mapping framework/estimator, and 
$P_M^i$ is the population estimator for the $i$-th electronic state 
(with $i=0,1,2$ for the $S_0$, $S_1$, and $S_2$ electronic states of ethylene, respectively).
Initial values for the electronic phase space variables are sampled using the focusing 
approximation~\cite{Bonella2003,Duke2015} such that the initial electronic state population 
is exactly 1 for the occupied $S_1$ state and 0 for the unoccupied states ($S_0$ and $S_2$).
This is achieved by sampling initial electronic mapping variables for 
the $i$-th electronic state from a circle with radius $x_i^2 + p_i^2=r_i^{2}$;
The radii for occupied and unoccupied states in each implementation are 
specified in Table~\ref{tab:samp_func_radii}, and are obtained by solving the corresponding
equations for the population estimators provided in 
Eq.~\ref{eq:el_seo_density} for the Wigner estimator in the MMST framework, 
in Eq.~\ref{eq:el_density} for the SC estimators in the MMST framework, 
and in Eq.~\ref{eq:3_elec_proj_cart} for $W$-representation in the SM framework.
Finally, in Eq.~\eqref{eq:lsc_pop_cf}, the electronic population estimator at time $t$, 
$P^i_M({\bf x}_t, {\bf p}_t)$, is evaluated using the time-evolved 
electronic positions and momenta obtained by propagating 
trajectory initial conditions under the classical analog 
symmetrized mapping Hamiltonian defined in Eq.~\eqref{eq:mmst_adiabH_sym}
for all three implementations.

%from Eq.~\ref{eq:wig_sampling} in the modified LSC-IVR and from Eq.~\ref{eq:spin_foc_sampling} in the spin LSC-IVR. 
%As briefly discussed in Sec.~\ref{sec:corr_spin}, in this manuscript we only consider the $W$-representation among the three SW transforms. 
%In case of the conventional LSC-IVR, solving Eq.~\ref{eq:seo_occ_sampling} for $F=3$ gives $x_{k}^{2}+p_{k}^{2} \approx 1.5589$ as the larger solution for the fully occupied state. 
%Hence, the sampling function is written as $\delta (x_{k}^{2}+p_{k}^{2} - 1.5589)$ for $k = S_1$ and $\delta (x_{j}^{2}+p_{j}^{2} - 1/2)$ for $j = S_0$ and $S_2$.

Classical equations of motion are integrated using the 
$4^\text{th}$ order Adams-Bashforth-Moulton predictor-corrector integrator with 
a time step of 1 a.u. time ($\sim 1/40$ fs) for a total time of $400$ fs.
The necessary classical forces at each time step are obtained from on-the-fly calculations of 
the adiabatic energies of the three electronic states, their gradients, and the nonadiabatic coupling vectors
using Pople's $6$-$31$+G$^{*}$ basis set and CASSCF(2o2e) with three states included in state averaging
in the electronic structure package GAMESS.\cite{GAMESS} 
%Trajectories are discarded if the maximum absolute deviation of the total energy exceeds $0.2$\% of the initial energy, which corresponds to the accessible thermal energy at $300$K for a system with approximate Franck-Condon excitation energy determined at CASSCF(2o2e) level ($\sim10$ eV). 

\section{\label{sec:result}Result and discussion}

\subsection{Photoisomerization}
\begin{figure}
%\captionsetup[subfigure]{position=t}
    \centering
    \includegraphics[width=8.5cm]{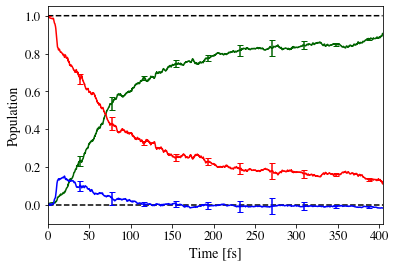}
    \caption{The population of the three electronic states obtained 
        from LSC-IVR simulation that employ the MMST mapping with 
        Wigner estimator are shown. The initial photo-excited $S_1$ state
        population is shown in red, the doubly excited $S_2$ state population 
        is shown in blue, and the ground $S_0$ state population is shown 
        in green. These results were obtained by averaging the number of the
    isomerization trajectories shown in Table~\ref{tab:traj_property}. 
    The error bars in the plot are obtained from the averages of $3$ subsets of 
trajectories, each containing $48$ trajectories.}
%\textcolor{red}{
%The other two methods do not have sufficient trajectory numbers to establish 
%statistically meaningful error bars.}}
    \label{fig:population}
\end{figure}

We discuss the results of our on-the-fly ab initio LSC study of ethylene photoisomerization
and dissociation in context with previous theoretical efforts using AIMS,~\cite{Martinez1998,Martinez2000,Mori2012,Allison2012} 
surface hopping,~\cite{Barbatti2005} and SQC~\cite{Frank2019,Frank2021} as well as experimental 
studies.~\cite{Satya1990,Balko1992,Lin1998,Lin2000,Farmanara1998,Farmanara1999,Stert2004,Lee2004,Lee2006,Mestdagh2000,Kosma2008,Karashima2022}
As with detailed earlier theoretical simulations of this system,~\cite{Martinez1998}
we include the ground state ($S_0$) and two excited states ($\pi \rightarrow \pi^* = S_1$ and $\pi^{*2} = S_2$), 
and exclude the low-lying Rydberg states that are expected to be present but appear
to play no significant role in the quenching process.~\cite{Martinez2000, Mori2012, Champ2016}
In FIG.~\ref{fig:population}, we plot the population correlation function for the three electronic states 
obtained using the Wigner population estimator in the MMST mapping framework.
We find that the initial photo-excited $S_1$ state decays to about 50\% of its original 
population on a 60-80 fs timescale, in good agreement with previous theory~\cite{Barbatti2005,Frank2019,Frank2021} and 
experiment.~\cite{Kosma2008} This is accompanied by a rapid but small population transfer to 
the higher lying $S_2$ state that, in turn, decays back to the ground state in about $100$ fs.

\begin{figure}
%\captionsetup[subfigure]{position=t}
    \centering
    \includegraphics[width=8.5cm]{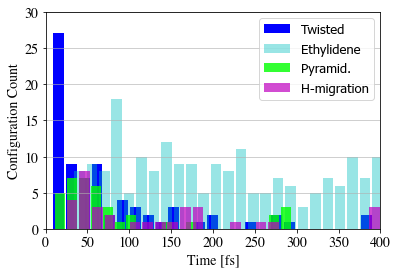}
    \caption{Characterization of molecular structures observed in $300$ trajectories from the conventional LSC-IVR simulation when the $S_{2}$-$S_{1}$ and $S_{1}$-$S_{0}$ energy gaps are small ($<0.2$eV). The observed structures can be classified into the four structures associated with the ethylene conical intersections (FIG.~\ref{fig:CI_structures}). The $S_{2}$-$S_{1}$ small energy gaps are characterized mostly by twisted structure and the $S_{1}$-$S_{0}$ gap by $\text{H}$-migration, pyramidalized or ethylidene structure.}
    \label{fig:small_Egap}
\end{figure}

\begin{figure}
\captionsetup[subfigure]{position=t}
    \centering
    \subcaptionbox{}{\includegraphics[width=3.0cm]{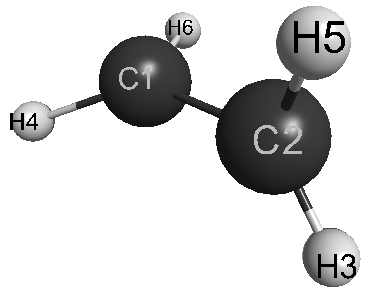}} \hspace{12mm}
    \subcaptionbox{}{\includegraphics[width=3.0cm]{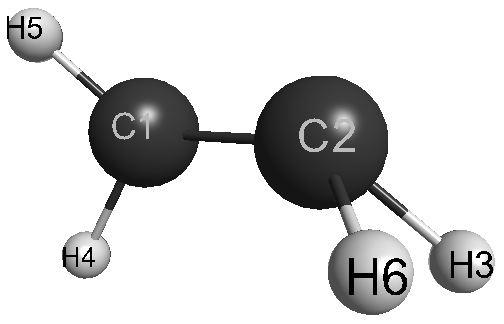}}\\ 
    \subcaptionbox{}{\includegraphics[width=3.0cm]{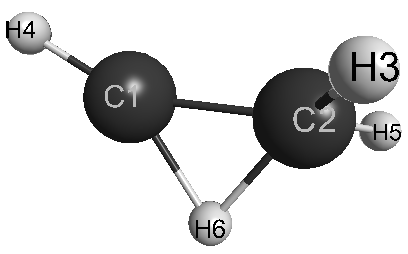}} \hspace{12mm}
    \subcaptionbox{}{\includegraphics[width=3.0cm]{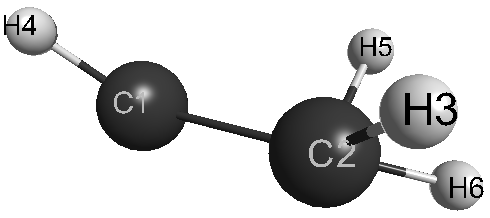}}
    \caption{The representative geometries of conical intersections encountered in the quenching of ethylene excited state through isomerization: (a) twisted, (b) pyramidalized, (c) H-migration, and (d) ethylidene.}
    \label{fig:CI_structures}
\end{figure}

Analysis of the trajectory ensemble employed in our LSC simulation yields insights into
the coupled nuclear motions that drive electronic state transitions. In FIG.~\ref{fig:small_Egap}, 
we show the timeline on which key ethylene molecular structures appear as trajectories
enter regions with a near-degeneracy between two electronic states, defined here as 
$E_{S1}-E_{S0} < 0.2$ eV and $E_{S2}-E_{S1} < 0.2$ eV. We find that there are four primary
structures, as suggested by previous work.~\cite{Martinez1998,Levine2007} 
These structures are shown in FIG~\ref{fig:CI_structures}: twisted, pyramidalized (pyramidalization of one or both $\text{CH}_2$ groups), 
H-migration, and ethylidene.  The twisted geometry dominates at early times ($\sim 10$ fs) weakening 
the $\pi$ electronic structure and enabling population transfer from the $S_1$ to the doubly 
excited $S_2$ state as seen in FIG.~\ref{fig:population}. The weakening of the $\pi$ bond also facilitates
the formation of the remaining three structures that occur at near-degeneracies between the $S_1$ and $S_0$ states.
Pyramidalization and H-migration appear at relatively early times ($<80$ fs) while
the formation of ethylidene occurs at later times and persists, in keeping with experiments 
that see signatures of this structure up to $600$ fs.~\cite{Allison2012}
%
%This sequence of structure dynamics is then followed by ethylidene formation, which keeps showing up for the rest of the dynamics and becomes the major molecular structure associated with small energy gaps between $S_1$ and $S_0$ states. 
%The result thus suggests that the initial quenching to the ground state is driven by the motion associated with pyramidalization and H atom migration in addition to ethylidene formation, and the long-term quenching occurs mainly through ethylidene formation. 
%It is experimentally shown that the presence of ethylidene persists up to $600$ fs,\cite{Allison2012} which is in agreement with what we see in FIG.~\ref{fig:small_Egap}.

%% RESULTS PART 1: Photoisomerization (show just the conventional LSC results)
%\subsection{\label{sec:isomerization}Photoisomerization}

\subsection{Photodissociation pathways}
In addition to isomerization, we observe a significant number of trajectories that describe photodissociation of ethylene upon excitation to $S_1$ as documented in Table~\ref{tab:traj_property}. 
This is in keeping with both previous experimental work~\cite{Satya1990,Balko1992,Lin1998,Lin2000,Lee2004,Lee2006,Kosma2008} 
and some theoretical simulations.~\cite{Chang1998,pena2002,Allison2012}

\begin{figure}
\captionsetup[subfigure]{position=t}
    \centering
    \subcaptionbox{}{\includegraphics[width=2.0cm]{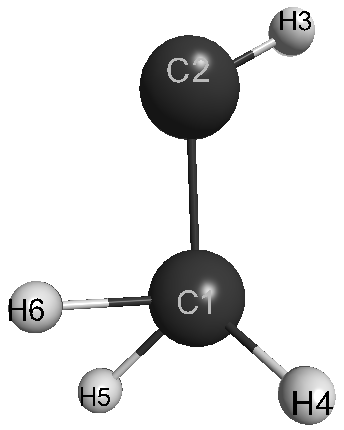}} \hspace{18mm}
    \subcaptionbox{}{\includegraphics[width=1.7cm]{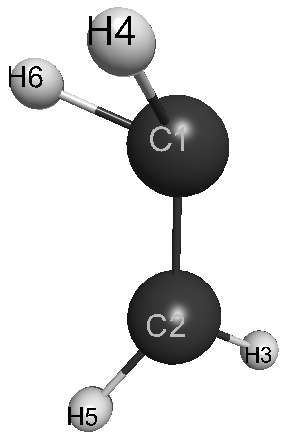}}
    \caption{The structures at transition states leading to $\text{H}_2$ elimination: (a) ethylidene resulting in a $\text{H}_2$ molecule and acetylene ($\text{HCCH}$) and (b) pyramidalized structure resulting in a $\text{H}_2$ molecule and vinylidene ($\text{H}_2\text{CC:}$).}
    \label{fig:H2_elim_TS}
\end{figure}

After photoisomerization, we see the elimination of molecular H$_2$ as the most significant channel
for photodissociation. In analyzing this subset of trajectories, we can identify the structures that 
are the major precursors of $\text{H}_2$ molecule production: about 39\% of the trajectories produce 
H$_2$ and acetylene ($\text{HCCH}$) from the ethylidene structure shown in FIG.~\ref{fig:H2_elim_TS}a, 
while the majority (52\%) produce H$_2$ and vinylidene ($\text{H}_{2}\text{CC:}$) from the pyramidalized geometry 
shown in FIG.~\ref{fig:H2_elim_TS}b.
These findings reproduce the two channels for H$_2$ molecule elimination
identified by experiments and the intermediate structures we report have been previously characterized 
by transition state calculations using electronic structure.~\cite{Lee2004,Lee2006}
%It has been experimentally shown that $\text{H}_2$ molecule elimination channel and $\text{H}$ atom elimination channel are accessible from $157$ to $193$ nm excitation, which corresponds to an excitation energy of $6.4$ to $7.9$ eV. 
%The breakdown of the simulated trajectories into isomerization with no dissociation, three types of dissociation ($\text{H}_2$ molecule elimination, $\text{H}$ atom elimination and C-C bond dissociation)  and energy conservation failure is summarized in TABLE~\ref{tab:traj_property}. 

%It has been experimentally indicated that there are four dissociation pathways of an excited ethylene molecule\cite{Lee2004,Lee2006}: $\text{H}_2\text{CCH} + \text{H}$, $\text{HCCH} + 2\text{H}$, $\text{HCCH} + \text{H}_2$ and $\text{H}_2\text{CC:} + \text{H}_2$. 
%Transition states leading to $\text{H}_2$ elimination from ethylene has been characterized by electronic structure calculation by Lee et al.\cite{Lee2006} 
%They identified two transition states, similar structures of which obtained from our conventional LSC-IVR simulation are shown in FIG.~\ref{fig:H2_elim_TS}. 
%The agreement of the major precursors of $\text{H}_2$ elimination obtained from our simulations with the transition states identified by the previous electronic structure calculation indicates that the $\text{H}_2$ elimination channel observed in LSC-IVRs is legitimate, not accidental.

A smaller but statistically significant number of our trajectories as reported in Table~\ref{tab:traj_property} also describe the elimination of 2 H atoms, the first leading to formation of the vinyl radical ($\text{H}_2\text{CCH}$), and the subsequent H atom elimination yielding acetylene. 
We note that while this mechanism is in keeping with experimental studies,~\cite{Lee2004,Lee2006} a very low but persistent yield (2\%) of the vinyl radical has been observed suggesting that it is possible to stop with the elimination of a single H-atom. 
We find $\sim 5\%$ of trajectories in our simulation correspond to the single H atom dissociation channel, which is in a reasonable agreement with this experimental observation. Although experimental measurements of the H$_2$:H branching ratio 
at $157$ nm excitation differ slightly, $0.44: 0.56$ in Ref.~\onlinecite{Lin2000} and $0.3 : 0.7$ in Ref.~\onlinecite{Lee2004}, both consistently 
find higher yields of $\text{H}$ atoms than $\text{H}_2$ molecule. The LSC-IVR simulations do not reproduce this 
ratio, yielding H$_{2}$:H $ = 0.66 : 0.34$, an error we attribute the level of electronic structure theory employed rather than the dynamics as discussed extensively in the context of C-C bond cleavage.

Finally, we note that in Table~\ref{tab:traj_property}, we show that 20\% of our trajectories 
result in unphysical C-C bond cleavage for which, to our knowledge, there is no experimental 
evidence other than in pump-probe or high energy ionization experiments,~\cite{Allison2012,Karashima2022} 
where the cleavage is explicitly targeted. It is important to identify the source of this error as 
due to either the nature of the dynamics employed or the underlying electronic structure. To unravel
this, we start by noting that the initial excitation energy in our simulation obtained with 
SA2-CASSCF(2o,2e) with $6$-$31+$G$^*$ basis, 
$\langle \Delta E_{S_{1}S_{0}}(t=0)\rangle$, is $9.77$ eV as reported in Table.~\ref{tab:traj_property}, 
a number significantly above the C-C bond dissociation energy of $7.7$eV at this level of theory.
Notably, our $\langle \Delta E_{S_{1}S_{0}}(t=0)\rangle$ is also significantly 
above the experimentally quantified Franck-Condon excitation energy of $7.6$ eV.\cite{Robin} 
%in our simulation $9.77\pm0.45$ eV, $9.78\pm0.47$ and $9.75\pm0.48$ for the conventional, modified and spin LSC-IVR, respectively, all being significantly higher than the experimental range. 
We find that using a more extensive basis (aug-cc-pVDZ) reduces the calculated excitation energy
to $8.86$eV, and further using second-order perturbative energy correction (XMCQDP) yields 
$7.63$eV, a value in close agreement with experiment. %$7.63$ eV from $\sim9.8$ eV, 
Unfortunately, at present, we cannot calculate the nonadiabatic coupling matrix element (NACME) in GAMESS
with the perturbation correction, so in order to test the dynamics we run an additional 50 trajectories
using the aug-cc-pVDZ basis set. We find that the number of trajectories that exhibit C-C bond cleavage
drops significantly from 20\% to only 10\%. This suggests that the underlying source of error leading
to unphysical trajectories in this case can indeed be attributed to the level of electronic structure theory.
The dependence of the nature of trajectories on the initial energy can also be seen by simply removing high energy trajectories: when we do not include trajectories with $\Delta E_{S_{1}S_{0}}(t=0) > 10$ eV, the fraction of trajectories exhibiting C-C cleavage drops by 10\%, while that of isomerization increases by 10\% and other categories remain almost unchanged. 
\subsection{Comparing the three different variants of LSC-IVR}
In discussing the results of our \textit{ab initio} study of photoisomerization and dissociation in 
ethylene, we confined ourselves to interpreting the results from the LSC simulation with 
MMST mapping and Wigner estimators. We now motivate this choice and further provide a detailed 
discussion of the three variants of LSC explored here, paying particular attention to energy conservation, observed reaction channels, and electronic population dynamics. 

\begin{figure}
    \centering
    \includegraphics[width=8.5cm]{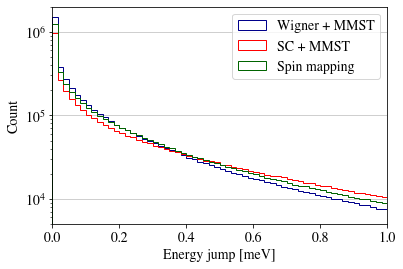}
    \caption{The absolute value of jumps in the total energy between two consecutive time steps in 300 simulated trajectories are compiled in a histogram for the three LSC variants: MMST mapping with Wigner estimator (blue), SC estimator (red), and spin mapping (green).}
    \label{fig:E_jump}
\end{figure}

Table \ref{tab:traj_property} summarizes the breakdown of the trajectory types obtained in the three different LSC implementations.  
We note that the energy conservation along a trajectory is generally considered poor in \textit{ab initio} implementations due to the self-consistent field calculations at every time step~\onlinecite{Frank2021}. 
For this reason, the effect of the initial conditions on the energy conservation of resulting trajectories can be masked by that of electronic structure calculations.
In FIG.~\ref{fig:E_jump}, we histogram the energy jump between consecutive time steps for all 300 trajectories generated in each LSC implementation. Recall that trajectory initial conditions for the nuclei are identical in all three implementations as is the Hamiltonian for dynamics; they only differ in the way the electronic mapping variables are sampled, and in the form of the population estimator employed at time $t$. Interestingly, FIG.~\ref{fig:E_jump} does capture the dependence on initial conditions \textemdash\, we find that 92\% of the trajectories in MMST mapping with Wigner estimator exhibit energy jumps of less than 1 meV, whereas this numbers drops to 82\% for the MMST with the Semiclassical estimator and to 88\% for spin mapping. 
All three simulations also yield fragmention trajectories that correspond to H$_2$ molecule elimination 
and H-atom dissociation, two channels that have been observed experimentally. 
All three also overestimate the likelihood of C-C bond dissociation but as discussed we believe this error can be attributed to electronic structure rather than the dynamics themselves.
%{\color{red}We note that since neither the AIMS nor surface hopping 
%employ ensembles of trajectories and do not, 
%typically report on the relative statistics of different pathways}, our discussion
%of accuracy is confined to comparison with experimental studies where possible.

\begin{table*} % page-wide table
\caption{\label{tab:traj_property} Results from the 3 variants of LSC-IVR are shown. 
Succesful trajectories are identified as those that undergo experimentally observed pathways - photoisomerization, $\text{H}_2$ elimination, and H atom elimination. 
C-C bond cleavage is only seen at extremely high excitation energies.  
The 1 "other" trajectory observed in the SC + MMST mapping and spin mapping
LSC implementations corresponds to an outlier trajectory that results in the molecule separating into individual atoms. 
We also report the average $S_1$ to $S_0$ excitation energy at $t=0$ for each subset of trajectories in eV as $\langle \Delta E_{S_{1}S_{0}} \rangle$.}
\begin{ruledtabular}
\begin{tabular}{lcccccc}
\multirow{2}{*}{LSC-IVR variants} & \multicolumn{2}{c}{Wigner + MMST mapping} & \multicolumn{2}{c}{SC + MMST mapping} & \multicolumn{2}{c}{Spin mapping}\\
\cline{2-3}\cline{4-5}\cline{6-7}\Tstrut
 & \text{Count} & $\langle \Delta E_{S_{1}S_{0}} \rangle$ & \text{Count} & $\langle \Delta E_{S_{1}S_{0}} \rangle$ & \text{Count} & 
 $\langle \Delta E_{S_{1}S_{0}} \rangle$ \\
\hline\Tstrut
%\textbf{Successful trajectories} &&&&&& \\
%\Tstrut\Tstrut
\text{Isomerization} & $144$ & $9.65 \pm 0.44$ & $47$ & $9.62 \pm 0.48$ & $73$ & $9.61 \pm 0.46$ \\
\Tstrut
$\text{H}_2$ elimination & $78$ & $9.78 \pm 0.40$ & $128$ & $9.75 \pm 0.42$ & $119$ & $9.75 \pm 0.51$ \\
\Tstrut
\text{H atom elimination} & $22$  & $9.76 \pm 0.46$ & $74$  & $9.82 \pm 0.47$ & $45$ & $9.71 \pm 0.47$ \Tstrut \\
%\hline\Tstrut
%\textbf{Failed trajectories} &&&&&& \\
\Tstrut\Tstrut
$\text{C-C cleavage}$ & $56$  & $10.06 \pm 0.41$ & $50$  & $9.93 \pm 0.54$ & $62$  & $9.97 \pm 0.38$ \Tstrut\\
%$\text{E conservation failure}$ & $10$ & $9.92 \pm 0.27$ & $55$  & $9.79 \pm 0.40$ & $25$  & $9.90 \pm 0.55$ \\
\Tstrut
$\text{Other}$ & $0$ & – & $1$  & – & $1$  & – \\[1mm]
\hline\Tstrut
$\text{Total}$ & $300$ & $9.77 \pm 0.45$ & $300$ & $9.78 \pm 0.47$ & $300$ & $9.75 \pm 0.48$ \\
\end{tabular}
\end{ruledtabular}
\end{table*}

\begin{figure}
    \centering
    \includegraphics[width=8.5cm]{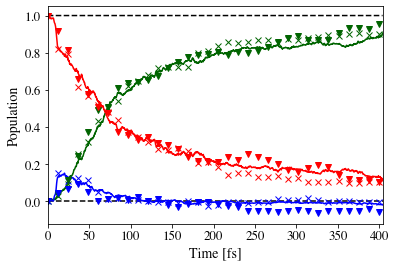}
    \caption{We compare the thermal population correlation functions obtained 
        from LSC simulations that employ the MMST mapping with 
        Wigner estimator ("$-$"), SC estimator ("$\blacktriangledown$"), and the spin mapping ("$\times$"). The initial photo-excited $S_1$ state
        population is shown in red, the doubly excited $S_2$ state population 
        is shown in blue, and the ground $S_0$ state population is shown 
        in green. These results were obtained by averaging the number of the
    isomerization trajectories shown in Table~\ref{tab:traj_property}.}
    \label{fig:pop_3_variants}
\end{figure}

All three LSC implementations exhibit qualitatively similar population dynamics and rates of quenching: 
the $S_2$ state is quickly populated $\sim10$ fs after photoexcitation, 
and it takes $60$ to $80$ fs for $50$\% of $S_1$ population to decay to the ground state, as shown in 
FIG.~\ref{fig:pop_3_variants}. In comparing the three approaches, we pay particular attention to the issue of 
negative electronic state populations: it is well known that 
the classical dynamics employed here to time evolve the electronic mapping variables preserves
the sum of the individual state populations, but does not constrain individual 
state populations to take on values between 0 and 1. Effectively, the 
classical dynamics employed here fails to properly constrain the mapping variables 
to the quantum mechanically allowed phase space.~\cite{Ananth2013} Although
individual trajectories might explore unphysical values of state population in all
three LSC implementations, we use the ensemble average populations shown in 
FIG.~\ref{fig:pop_3_variants} to identify the `best' choice; it is clear that 
LSC with MMST mapping and Wigner populations as well as spin mapping LSC yield ensemble
average state populations that are between 0 and 1, whereas the 
SC estimator in the MMST mapping framework yields a significant negative value 
for the $S_2$ state population. 

Based on previous studies, we expected spin mapping LSC to significantly outperform the 
MMST mapping approaches, but the trajectory breakdown in Table~\ref{tab:traj_property}
and energy conservation shown in FIG.~\ref{fig:E_jump} make it clear that is not necessarily the case. We find that both in terms of energy conservation
and the overall number of C-C bond cleavage trajectories, MMST with Wigner estimator 
emerges as the better implementation. 
We further note that careful analysis of energy conservation in all three implementations 
yields no correlation between the frequency with which individual trajectories exhibit negative populations and trajectories that fail to conserve energy. While
this is likely something that should be analyzed on a case-by-case basis, 
this finding does provide some reassurance that an individual trajectory
exhibiting negative electronic state population at a given time 
does not always lead to unphysical behavior in terms of the overall system dynamics.

\section{Conclusion}
We make the case that employing \textit{ab initio} classical trajectories 
within the semiclassical LSC mapping framework shows great promise 
as an efficient and accurate on-the-fly simulation technique for 
the study of nonadiabatic processes. 
In the specific case of ethylene, we show that the use of an ensemble of trajectories in LSC allows us to directly calculate correlation functions, and to identify different, statistically significant reaction pathways with support from previous experimental observations and theoretical simulations.
%, something {\color{red}that is not possible using wavefunction based methods like AIMS or individual trajectory methods like surface hopping.}

We also explore three different implementations of LSC that differ
in the initial conditions used for the electronic mapping variables
and the electronic state population estimators. We establish that 
although all three yield relatively similar population dynamics, MMST mapping with Wigner estimators emerges as the best choice in this study on the basis of energy conservation, positive ensemble average electronic populations, and the relatively small number of C-C bond cleavage trajectories.
Given previous studies highlighting the favorable properties of spin mapping, 
however, further case studies must be made before this observation can be generalized.

Finally, we discuss the limitations of this approach at present. 
Like all \textit{ab initio} dynamics, it is clear that the level of electronic
structure plays a significant role in the overall accuracy 
of our findings. Setting that aside, we note that while the LSC 
implementation is streamlined and involves no free parameters such 
as hopping probabilities, decoherence corrections, or spawning 
thresholds, there are a few advances that would allow us to move
towards even more efficient implementation. Most notably, the 
development of improved integrators that will allow us to implement dynamics 
under the symmetrized Hamiltonian with a larger time step, and 
a more rigorous study of the dependence of mapping variable dynamics
on the initial conditions to minimize the number of trajectories that 
yield negative electronic state populations.

\begin{acknowledgments}
    The authors acknowledge funding through NSF CAREER Award No. CHE-1555205. The authors thank Prof. Ben Levine for helpful discussions about the theory and implementation of AIMS.
\end{acknowledgments}

\nocite{*}
\bibliography{manuscript}% Produces the bibliography via BibTeX.

\end{document}